\documentclass[conference]{IEEEtran}
\IEEEoverridecommandlockouts
\usepackage[T1]{fontenc}
\usepackage[utf8x]{inputenc}
\usepackage{cite}
\usepackage{amsmath,amssymb,amsfonts}
\usepackage{algorithmic}
\usepackage{graphicx}
\usepackage{textcomp}
\usepackage{xcolor}
\usepackage{balance}
\usepackage{xspace}
\usepackage{multirow}
\usepackage{float}
\usepackage{soul}
\usepackage{microtype}
\usepackage{url}
\usepackage{booktabs}
\usepackage{soul}

\def\BibTeX{{\rm B\kern-.05em{\sc i\kern-.025em b}\kern-.08em
    T\kern-.1667em\lower.7ex\hbox{E}\kern-.125emX}}

\newboolean{showcomments}
\setboolean{showcomments}{true}         
\ifthenelse{\boolean{showcomments}}
{\newcommand{\nb}[2]{
		\fbox{\bfseries\sffamily\scriptsize#1}
		{\sf\small$\blacktriangleright$\textit{\textcolor{red}{#2}}$\blacktriangleleft$}
	}
}
{\newcommand{\nb}[2]{}
	
}

\newcommand{\toolname}{\textsc{VARYS}\xspace}

\begin{document}

\title{Anomaly Detection As-a-Service}

\author{\IEEEauthorblockN{Marco Mobilio, Matteo Orr\`u, Oliviero Riganelli, Alessandro Tundo, Leonardo Mariani}
	\IEEEauthorblockA{\textit{University of Milano - Bicocca} \\
		Milan, Italy \\
		\{marco.mobilio $|$ matteo.orru $|$ oliviero.riganelli $|$ alessandro.tundo $|$ leonardo.mariani\}@unimib.it}}


\maketitle

\begin{abstract}
Cloud systems are complex, large, and dynamic systems whose behavior must be continuously analyzed to timely detect misbehaviors and failures. Although there are solutions to flexibly monitor cloud systems, cost-effectively controlling the anomaly detection logic is still a challenge. In particular, cloud operators may need to quickly change the types of detected anomalies and the scope of anomaly detection, for instance based on observations. This kind of intervention still consists of a largely manual and inefficient ad-hoc effort.

In this paper, we present Anomaly Detection as-a-Service (ADaaS), which uses the same as-a-service paradigm often exploited in cloud systems to declarative control the anomaly detection logic. Operators can use ADaaS to specify the set of indicators that must be analyzed and the types of anomalies that must be detected, without having to address any operational aspect. Early results with lightweight detectors show that the presented approach is a promising solution to deliver better control of the anomaly detection logic.
\end{abstract}


\section{Introduction}
\label{sec:introduction}

The adoption of cloud computing technologies can facilitate achieving sophisticated capabilities, such as scalability, dynamic resource allocation, and fault tolerance~\cite{yazir_dynamic_2010,josep2010view,buyya2009cloud,vaquero_dynamically_2011}.
However, it is not easy to design applications that behave properly in highly dynamic cloud environments~\cite{sabahi_cloud_2011,orru2018chmmiot}, and their health and reliability must be constantly monitored to detect possible anomalous behaviours~\cite{aceto2013monitoringsurvey}. 

There are several monitoring frameworks that can be used to deploy probes that collect data from cloud infrastructures (e.g., the CPU consumption of a machine), platforms (e.g., the bandwidth consumption of a specific service), and  applications (e.g., the number of users connected to an application running on the Cloud)~\cite{amazon_2019_cloudwatch, calero_monpaas:_2015, hp_2019_monasca}. Since configuring a monitoring framework, and reconfiguring it after every change that may occur in the monitored system, might be challenging, it is possible to use Monitoring-as-a-Service (MaaS) solutions, which support the declarative (re-)configuration of the monitoring system~\cite{shatnawi2018chmm,calero_monpaas:_2015, hp_2019_monasca}. That is, the operator specifies the monitoring objective and the MaaS solution  automatically takes all the actions necessary to collect and present the requested data to the operator.


Monitoring a software system is just a prerequisite to reveal misbehaviours because operators still have to inspect the collected data, to identify any problem that may require their intervention. Since the amount of collected data can be huge and it is not feasible to assume operators can be constantly looking at the data, the detection of anomalous behaviors must be automated. 

Anomalies can be detected in many different ways using different strategies and indicators, and also the anomaly detection requirements may change easily when the collected data change, it is thus important to be able to flexibly work with the anomaly detection logic. In particular, operators should be able to quickly modify the data and the strategy used to detect anomalies (e.g., detecting anomalies based on threshold values or detecting anomalies based on historical data). To this end, this paper early investigates the idea of exploiting the as-a-service paradigm also in the context of anomaly detection, delivering Anomaly Detection as-a-Service (ADaaS). The main principle is that the operator must be able to specify declaratively the kind of anomalies that must be detected (e.g., threshold- or peak-based) and the data that must be used to detect anomalies (e.g., an indicator about the memory or the CPU), while the technical infrastructure does the work of deploying and putting into operation the requested anomaly detection framework automatically.

This paper  presents our work about delivering ADaaS for cloud systems. We defined our approach on top of VARYS~\cite{tundo:2019}, which is a technology-agnostic MaaS framework for cloud systems. ADaaS relies on an architecture that is designed to dynamically and automatically deploy, redeploy, and un-deploy anomaly detectors, based on operator's needs.

The paper is organized as follows. Section~\ref{sec:maas} describes the VARYS framework. Section~\ref{sec:adaas} presents how we extended VARYS to achieve ADaaS. Section~\ref{sec:evaluation} presents a preliminary evaluation of anomaly detection. Section~\ref{sec:related} discusses related work. Finally, Section~\ref{sec:conclusion} provides final remarks.


\section{VARYS Framework}
\label{sec:maas}

%

\toolname is a \emph{technology-agnostic} Monitoring-as-a-Service solution that can address KPI monitoring at \emph{all levels} of the cloud stack. The main purpose of \toolname is to allow users to manage their monitoring systems through a declarative approach, without having to deal with the underlying technologies, whose management is encapsulated in specific components of the architecture. 

\toolname is \emph{model-driven}, that is, the monitoring goals are selected from a tree-like model where nodes represent quality attributes, which are decomposed into finer-grained quality attributes, until reaching the leafs of the tree that represent measurable properties. Operators can select monitoring goals at any level of the tree. A selected monitoring goal is then mapped to its measurable properties and the corresponding probes are automatically deployed. Metadata associated with probes and target services allow \toolname to automatically deal with technical choices, without bothering the operator.




\toolname is also \emph{reconfigurable} since it can be used to deploy and undeploy probes based on the actual needs of the operator. 


The architecture of \toolname is \emph{flexible} enough to be able to accommodate changes in the tree-like model, in the set of probes, and in the underlying technologies used to implement data collection. For instance, VARYS can interchangeably use Elasticsearch~\cite{elastic_2019_elasticsearch} and Prometheus~\cite{prometheus} as monitoring technologies.

Compared to other MaaS frameworks~\cite{amazon_2019_cloudwatch, calero_monpaas:_2015, hp_2019_monasca}, VARYS is less tailored to specific technologies or platforms and it is not limited to a set of KPIs, for instance low-level KPIs such as memory consumption or network bandwidth. It also exposes high-level APIs to express the monitoring intents declaratively, reducing the integration barriers.

Note that we started from \toolname~\cite{tundo:2019} to define ADaaS because \toolname provides a flexible API that facilitates integration. However, the ADaaS principle is not limited to \toolname and can be developed also using other MaaS frameworks. 


\section{Anomaly Detection as a Service}
\label{sec:adaas}

\subsection{The Approach}
ADaaS delivers the capability to define both the set of anomaly detection strategies and the indicators that should be checked with these strategies, declaratively. For example, an operator may use ADaaS to automatically deploy a threshold-based anomaly detector for CPU consumption of some containers and a history-based anomaly detector for memory consumption of a different set of containers.

To work properly, ADaaS has to integrate with a monitoring framework that collects the indicators about the behavior of the target cloud system in a time-series database, which is accessed by ADaaS to compute anomalies. This means that ADaaS can be used with regular monitoring frameworks, but it is more effective when used jointly with a MaaS framework, because both the collected KPIs and the anomaly detection strategies could be changed dynamically. In particular, our prototype has been designed to seamless integrate with VARYS~\cite{tundo:2019}. 

\subsection{Architecture}
\begin{figure}[hbt]
  \centering
  \includegraphics[width=0.5\textwidth]{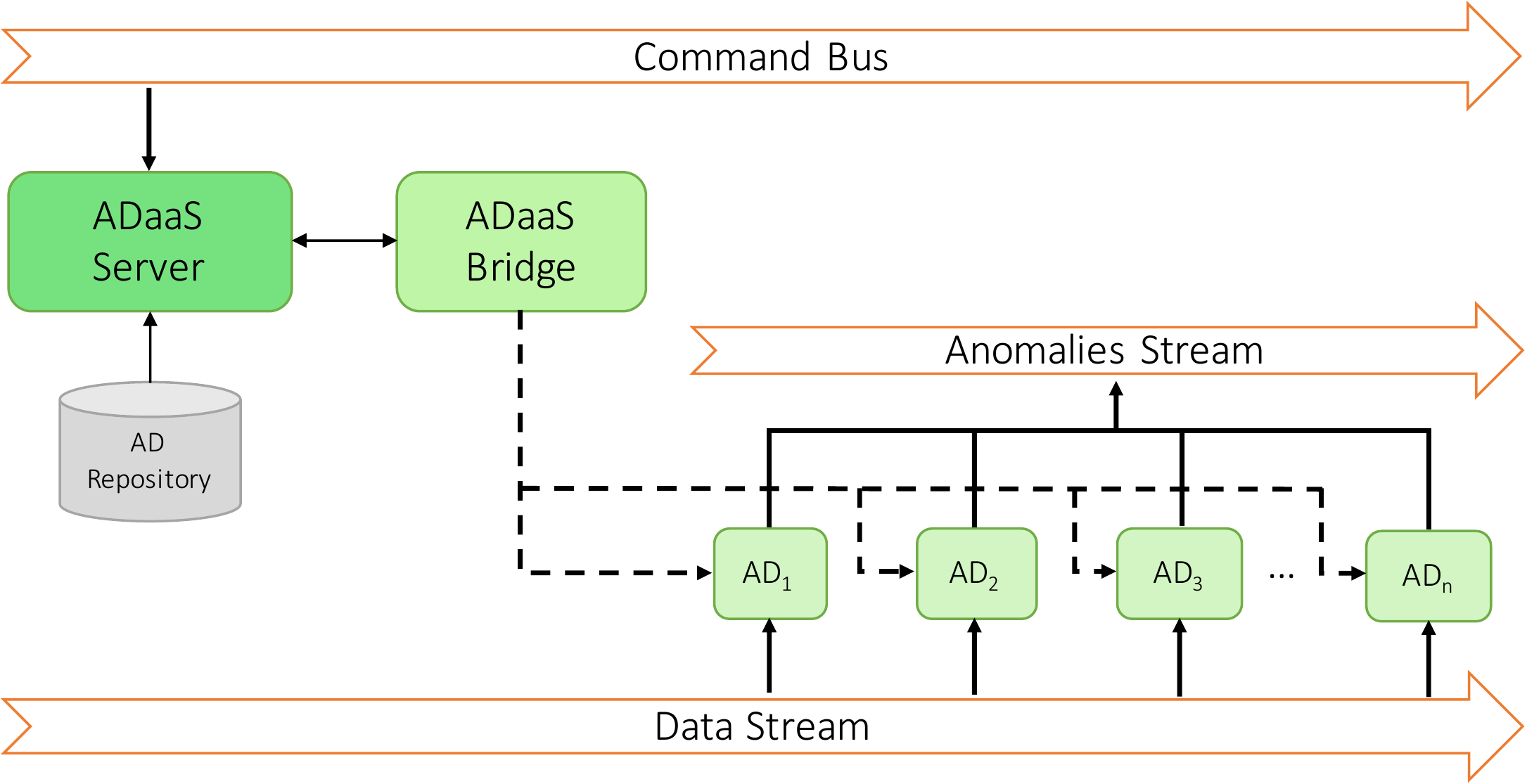}
  \caption{The ADaaS Architecture.}
  \label{fig:adaas_architecture}
\end{figure}

The architecture of the ADaaS framework is shown in \figurename~\ref{fig:adaas_architecture}.

The \textit{ADaaS Server} receives commands from the \emph{command bus} and maps these commands into the concrete operations that must be performed to actuate them. In particular, the received commands take the form of pairs $\langle \textit{KPIs}, \textit{analysis} \rangle$, where $\textit{KPIs}$ represent the set of indicators that must be analyzed, and $analysis$ represents the anomaly detection strategy that must be used. The ADaaS Server exploits a repository of anomaly detectors to select the ones that correspond to the requested analysis (the detectors in the repository are augmented with metadata to make the selection possible), and requests the ADaaS Bridge to deploy them.



The \textit{ADaaS Bridge} is the component in charge of deploying the anomaly detectors as requested by the ADaaS Server while taking the target technology under consideration. That is, the (technology-agnostic) logic for identifying the detectors to be deployed is in the ADaaS Server, while the bridge exploits information about the cloud platform to make optimal deployment of detectors. In particular, it provides CRUD and status operations that allow the ADaaS Server to know if a detector has been fully deployed and is operational, to know the list of active detectors, and more in general to manage the lifecycle of the deployed detectors. 


The \textit{Anomaly Detectors} $\textit{AD}_1 \ldots \textit{AD}_n$ are the actually deployed detectors. The architecture is intended to favour the usage of (potentially many) small, lightweight anomaly detectors, each one targeting a different kind of anomaly on different indicators, instead of using heavy monolithic and holistic machine-learning components. Specifically, small detectors, fast to deploy/un-deploy, that require small or no learning time and that focus on a very limited set of KPIs (often just one) are a natural choice for situations in which the monitored KPIs change frequently, as in dynamic cloud environments. 
Since anomaly detectors read and write data from time-series databases, the anomaly detectors can be combined: the anomalies produced by a detector can be the input to another detector to generate higher-level anomalies from lower-level anomalies.

One bus and two data streams are also part of the architecture in \figurename~\ref{fig:adaas_architecture}. The \textit{Command bus} is used to send commands to the ADaaS Sever. The \emph{Data Stream} and \emph{Anomaly Stream} are input and output connections, respectively. These connections can be linked to actual streams or time series databases (using a single time-series database for both input and output operations is indeed possible). 

\subsection{Scenario: Deployment of the Anomaly Detectors}

In this section, we summarize the behavior of ADaaS when a request for changing the set of deployed anomaly detectors (to add/remove/replace detectors) must be served. ADaaS works according to the following sequence of steps:
\begin{enumerate}
	\item The ADaaS Server receives a request from the command bus.
	\item The ADaaS Server compares the content of the request and the status of the deployment to identify the set of changes that must be performed on the existing detectors. The status of the deployed detectors is obtained by querying the ADaaS Bridge. The metadata associated with the available probes (e.g., the name of the specific KPI to query on the time-series database) enable the completion of this step.
	\item The ADaaS Server sends a specific request for changes to the ADaaS Bridge.
	\item The ADaaS Bridge maps the request into a set of concrete operations that depend on the target technology, such as the type of cloud platform used (e.g., based on containers or virtual machines) and the type of time-series database used.
	\item The ADaaS Bridge monitors the status of the deployment of the detectors and makes it available to the ADaaS Server, and to dashboards that might be used to visualize the status of the system.
\end{enumerate}

 
 
 \subsection{Anomaly Detectors}
Although a virtually unlimited number of  anomaly detectors can be defined and deployed in the context of ADaaS, we started with the following three general anomaly detectors that can be applied to any numeric indicator:
\begin{itemize}
	\item \emph{Fixed Threshold Anomaly Detector}: This anomaly detector depends on the value of parameters $V$ and $n$, which are specified in the request received by the ADaaS Server and are set when the detector is first deployed. The anomaly detector works as follows. If $n$ consecutive values in the checked indicator have values above threshold $V$, the anomaly detector fires the anomaly. More formally, given a time series $\ldots, v_{t+1},\ldots v_{t+n}$ of values, if $v_{t+i} \geq V, \forall i=1,\ldots n$, this detector fires an anomaly at time $t+n$, otherwise no anomaly is fired.	
	\item \emph{Sigma Limit Anomaly Detector}: This anomaly detector can detect burst, outliers, and short term variations in the analyzed indicators. It works by comparing the current values to the mean and variance of the values collected in the recent past. If the current value can be classified as an outlier, an anomaly is fired. In particular, a value $v_{t+1}$ is reported as anomalous if equation~\ref{eq:sigma} holds
	 \begin{equation}	\label{eq:sigma}
 	d(v_{t+1}, mean(v_{t-\Delta},v_t)) > \sigma *std(v_{t-\Delta},v_t)
 	\end{equation}
	 where:
	 \begin{itemize}
	 	\item $d(x,y)$ represents the distance between two values. In many cases this can be a simple difference, however more complex distance functions can be used for instance when samples are tuples.
	 	\item $v_{t+1}$ is the value that is checked and $t+1$ is the time at which the anomaly can be fired.
	 	\item $\Delta$ is the size of the window that contains the values from $t-\Delta$ to $t$.
	 	\item $mean(v_{t-\Delta}, v_t)$ is the mean of all values in the window.
	 	\item $std(v_{t-\Delta},v_t)$ is the standard deviation of all values in the window.
	 	\item $\sigma$ is a parametric multiplier that determines the sensitivity of the detector (usually $\sigma=3$ is used).
	\end{itemize}
	\item \emph{Mean Shift Anomaly Deetector}: This anomaly detector can detect long-term changes in KPIs. It is aimed at indicators (or combinations thereof) that are supposed to remain stable in normal usages scenario (e.g. memory consumption, or the ratio between memory consumption and number of requests being served). Formally, the detector fires if equation~\ref{eq:meanShift} is valid
	\begin{equation} \label{eq:meanShift}
		|mean(v_{t-2\Delta-1}, v_{t-\Delta-1}) - mean(v_{t-\Delta}, v_{t})| > \lambda
	\end{equation}
	where:
	\begin{itemize}
	\item $\Delta$ is the chosen size for the window.
	\item $mean(v_{t-\Delta}, v_{t})$ is the mean of all the last $\Delta$ values.
	\item $mean(v_{t-2\Delta-1}, v_{t-\Delta-1})$ is the mean of the previous $\Delta$ values.
	\item $\lambda$ is a parameter that determines the sensitivity of the detector. As an example, we may set $\lambda = std(v_{t-2\Delta-1}, v_{t-\Delta-1})$, so that the detector fires an anomaly if the difference between the mean of the current window and the mean of the previous window is greater than the standard deviation of the previous window.
	\end{itemize}	
\end{itemize}

\subsection{Design of the Anomaly Detectors}

\begin{figure}[hbt]
  \centering
  \includegraphics[width=0.5\textwidth]{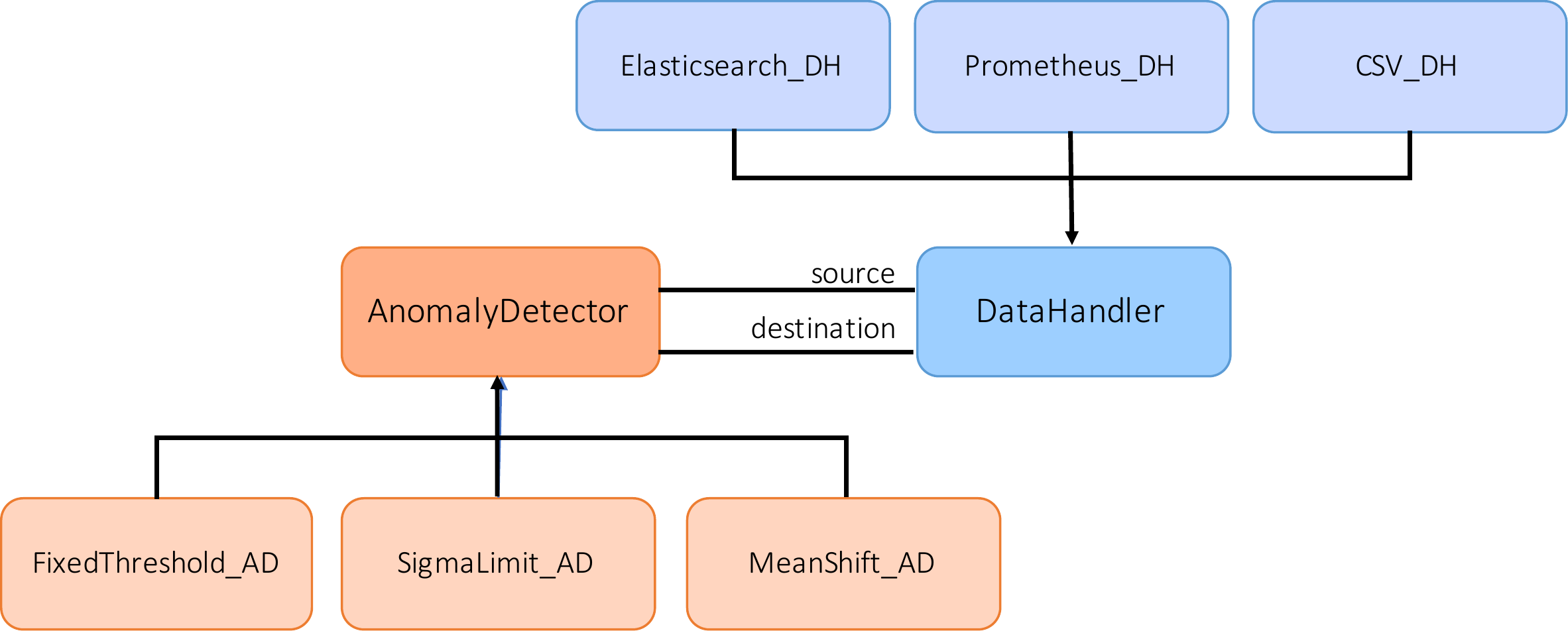}
  \caption{Design of the anomaly detection and data handling logic.}
  \label{fig:ad_components}
\end{figure}

To flexibly support the various combinations of time-series databases, and more in general the various sources of data, we separated the data  handling logic from the anomaly detection logic, using the design shown in \figurename~\ref{fig:ad_components}.

In particular, the specific anomaly detectors are extensions of the general class \textit{AnomalyDetector}, which implements the behavior that is in common to every anomaly detector. This includes:  
\begin{itemize}
	\item The instantiation of the source \textit{DataHandler} that connects the detector to the source time-series database with the indicator(s) that must be checked.
	\item The instantiation of the target \textit{DataHandler} that connects the detector to the target time-series database to write an indicator with the sequence of anomalies that have been fired.
	\item The capability to run the anomaly detection logic with the appropriate frequency.
\end{itemize} 

To implement new anomaly detectors, developers only need to create a subclass of \textit{AnomalyDetector} with the proper anomaly detection algorithm. 

The implementation of the data handlers follows a similar structure, with the general data handling logic implemented in the \textit{DataHandler} class and the specific operations to read and write from different kinds of sources implemented in the subclasses. We also defined a common output format for all the data handlers. Right now we support data handlers that can read and write from Elasticsearch, Prometheus, and CSV files. While the first two data handlers are designed to support online analysis, the third data handler has been implemented to support repeatable offline data analysis. As an example, the \textit{Elasticsearch\_DH} queries Elasticsearch for new documents with the specific sampling rate and query specified in the config file. The \textit{CSV\_DH} has been implemented specifically for feeding offline experimental data, thus allowing to perform the same analysis on different data without the need to modify the code of the detector.

\subsection{Implementation Details}
Our current prototype implementation supports the deployment of anomaly detectors based on containers and their dynamic configuration based on shared volumes. Both the ADaaS Server and the ADaaS Bridge are Python components that can be queried with a REST API. The individual anomaly detectors are also implemented as Python components, although our architecture does not impose any constraint on the language to be used to implement the detectors.

\section{Evaluation}
\label{sec:evaluation}

In this section, we report early results obtained by applying failure prediction in combination with the sigma limit anomaly detection.

\subsection{Subject System}
The objective of the evaluation is to assess the approach with a realistic case, to observe our prototype in operation and determine to what extent anomaly detection may support failure prediction capabilities. To this end, we selected Clearwater \cite{clearwater} as subject system, which is an open source implementation of an IP Multimedia Subsystem. A standard installation of Clearwater consists of six components running in separated VMs.


In this evaluation, we detect anomalies from 315 KPIs collected from all the components of the system and from different layers. Examples of KPIs are the number of rejected requests, the average latency, and the used memory. All the detectors have been deployed declaratively.

\subsection{Experiment Setup}
We evaluate the quality of the anomaly detection mechanism by studying how it can support the automatic prediction of failures. In particular, we execute Clearwater without producing any failure and we train a one class Support Vector Machine~\cite{chang2011libsvm} for predicting failures based on the reported anomalies. Note that a one class Support Vector Machine can be trained from positive samples only and also failure-free executions produce anomalies. 

More specifically, we use a sliding window of fixed length to generate multiple windows with anomalies, starting from the anomalies reported while running the system with normal (non-failing) executions. The trained model is able to predict failures based on the degree of difference between the anomalies produced by the actual observed behavior and the sets of anomalies observed in the training phase.

To generate the normal executions to train the model, we designed a workload that implements a business scenario which performs calls based on calendar patterns such as time of the day and day of the week. In particular, we consider regular office hours and our workload variation patterns are of two types: daily variations and hourly variations. In the former case, the system is busier on working days of the week (Monday to Friday). In the latter case, our system experiences an increase in calls during the daily hours with peaks around 2pm, and only a few calls during the night.

Predictions are generated by checking the anomalies within sliding windows of the same length of the windows used in the training phase. If the trained failure predictor consistently returns a failure prediction for several windows, the failure prediction is finally returned to the operator. We determined empirically that collecting a sample per minute and reporting a failure prediction when the prediction is stable for 7 samples works quite well.

To evaluate failure prediction in presence of faults, we injected the following four types of faults: packet loss faults, excessive workload conditions, memory leaks, and CPU hogs. We injected a fault per type: a packet loss fault in Sprout's\footnote{Sprout (SIP Router) acts as a combined SIP registrar and authoritative routing proxy, and handles authentication.} virtual network interface; a memory leak in Bono's\footnote{Bono (Edge Proxy) provides the entry point for the client connections to the Clearwater system.} VM; a CPU hog in Sprout's VM; and an excessive workload for the entire Clearwater system. We activated the injected faults according to three activation patterns: (1) the fault is activated with a same frequency over time; (2) the fault is activated with a frequency that increases exponentially, resulting in a shorter time to failure; (3) the fault is activated randomly over time. In total we collected results for 12 experiments with failing executions (4 fault types time 3 activation patters) and 1 experiment in which the system has been executed in its normal operating conditions without faults or abnormal workloads.

\subsection{Results}
Since the failure prediction ability depends on both the length of the sliding window and the length of the history used by the sigma limit anomaly detector to compute thresholds, we started by empirically studying the impact of these two parameters on the effectiveness of the technique. Note that the sliding windows should be big enough to contain a sufficient amount of information to predict failures and small enough to be practical and sensitive to failure symptoms. In the evaluation, we studied sliding windows of size 15, 20 and 25 minutes (\emph{window size}). In terms of historical data used to train the anomaly detector, we considered to use the values of the indicators observed in the last 10, 20, 50 and 100 minutes (\emph{past data interval}).

\begin{figure}[hbt]
  \centering
  \includegraphics[width=0.45\textwidth]{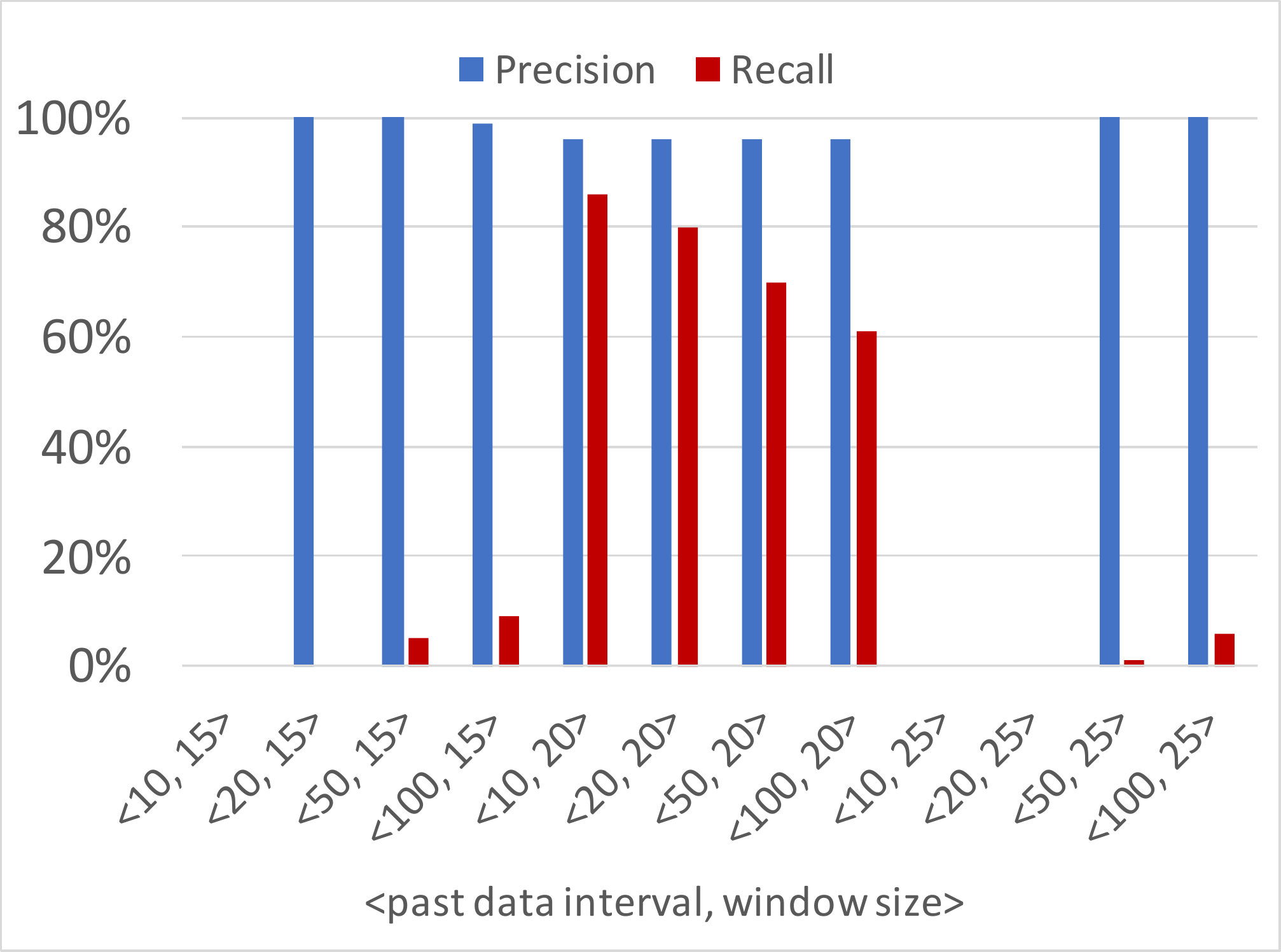}
  \caption{Precision and Recall.}
  \label{fig:precision}
\end{figure}

Figure~\ref{fig:precision} shows the results obtained for the standard measures of precision and recall: Precision is the ratio of correctly predicted failures over all predicted failures; Recall is the ratio of correctly predicted failures over actual failures. Values are averages over all the experiments.

Results indicate that the values of the studied parameters impact on predictions, and that a sliding window of 20 minutes and detectors working on the data reported in the last 10 minutes reach the best prediction effectiveness among the experimented sizes in term of precision (96\%) and recall (86\%). It is interesting that the resulting approach seldom generates false positives (96\% precision) and predicts failures for the vast majority of the samples that have been analyzed (86\%). Note that 86\% recall implies that for 86\% of the windows collected when a failure occurs, the failure is actually predicted. Thus failure predictions have been reported for all the analyzed failures. 

Since it is important to know how early failures can be predicted, Table~\ref{tbl:results} shows the average time occurring between a failure prediction and the failure occurrence (i.e., a system crash or the system being unable to serve calls). Failures are predicted well in advance, giving to operators the opportunity to intervene. 


\begin{table}[ht]
\centering
\caption{Results for each Fault Type.}
\label{tbl:results}
\begin{tabular}{@{}lc@{}}
\toprule
\textbf{Fault Type} & \textbf{Failure Prediction Lead-Time (avg)} \\ \midrule
CPU hog & \textgreater 12 hours \\
Memory Leak & 150 mins \\
Packet Loss & 136 mins \\
Excessive Workload & 161 mins \\ \bottomrule
\end{tabular}
\end{table}

Finally, we compared the effectiveness of the failure prediction based on sigma limit anomaly detectors, which is a lightweight detector, to the effectiveness of Hierarchical Temporal Memory (HTM)\cite{AHMAD2017134}, which exploits models that replicate the structural and algorithmic properties of the neocortex to identify anomalies in an unsupervised environment. The HTM models have been trained on the same data used for the lightweight anomaly detectors. Table~\ref{tbl:fpComparison} reports the precision, recall, and memory usage of the two methods. We can observe that the effectiveness of the lightweight anomaly detectors used in ADaaS and the effectiveness of HTM are similar, but the sigma limit method reduced memory usage by 72\%. This result confirms our initial intuition that large cloud systems with thousands or millions of running services can be more effectively addressed with scalable lightweight anomaly detectors.

\begin{table}[t]
\centering
\caption{Failure Prediction Comparison.}
\label{tbl:fpComparison}
\begin{tabular}{@{}lccc@{}}
\toprule
\textbf{Anomaly Detection Method} & \textbf{Precision} & \textbf{Recall} & \textbf{Memory Usage} \\ \midrule
Sigma Limit & 96\% & 86\% & 70 MB \\
HTM & 100\% & 89\% & 250 MB \\ \bottomrule
\end{tabular}
\end{table}

This result is preliminary and further experiments are necessary to generalize the collected evidence.


\section{Related Work}
\label{sec:related}


The body of knowledge related to anomaly detection is large since it covers a number of disciplines (such as, data science, machine learning, and statistics) and includes a wide variety of techniques (such as, classification and clustering techniques) spanning over multiple applicative domains (such as, intrusion and fraud detection, and medical informatics)~\cite{chandola_anomaly_2009}. There might be a large variety in the approaches but, by and large, the underlying rationale is looking for patterns of either features or behaviours that differ from what is considered normal or regular~\cite{Chandola:2008,chandola_anomaly_2009}. 
 
%

A large variety of anomaly detection techniques have been designed to specifically target  
data collected sequentially, such as in the case of \emph{time series}. 

However, many anomaly detection techniques simply process data in batches and for this reason are unsuitable for real-time streaming applications. This is the case of Symbolic Aggregate Approximation~\cite{keogh:2005}, which has been used to find the most unusual subsequences within a time series, the supervised learning approach by Hermine et al.~\cite{AKOUEMO2016948}, which leverages a combination of a Bayesian maximum likelihood classifier and a linear regression model to spot anomalies in temporal structures, 
and the Netflix's robust principle component analysis (RPCA) method~\cite{surus:2015}, and Yahoo's EGADS~\cite{Laptev:2015:GSF:2783258.2788611}.

The spread of the cloud computing technologies~\cite{gartner_forecast_21.4}, with their inherent dynamic nature, which favours adaptability, extendability, and scalability of resources and users, led to the emerging popularity of streaming applications, which are characterized by the processing of a continuous sequence of data in real-time. 
In the field of cloud computing, various techniques for detecting anomalies have been adopted and redefined~\cite{Tan:anomalyPrediction:ICDCS:2012,Dean:UBL:ICAC:2012,Sauvanaud:Anomaly:ISSRE:2016,Mariani:ICST:18}. However, common usage scenarios for cloud administrators and practitioners do not fit well with these approaches that use complex statistical analysis, are not scalable, and often require supervised learning of training sets coming from a historical data. 
On the contrary, there is a demand for anomaly detection strategies able to work online, in a continous fashion without or with a minimal training, with no need of storing the entire stream, that can adapt dynamically to the environment, producing alerts and returning anomalies as fast as possible~\cite{AHMAD2017134,Chandola:2008}.
 

Semi-online anomaly detection solutions partially respond to these requirements because they offer online anomaly detection abilities after an initial learning phase~\cite{Chen:2015,Spinosa:2007}. For instance, OLINDA identifies anomalies based on the concepts of novelty behaviour and drift~\cite{Spinosa:2007} and the approach by Chen et al. provides anomaly detection capabilities in the context of cyber-physical systems~\cite{Chen:2015}. 

Recently Subutai et al. introduced a novel approach for the anomaly detection of streaming data based on Hierarchical Temporal Memory (HTM)~\cite{AHMAD2017134}, a model able to capture interesting features of time-series data. This technique is unsupervised but the computational burden, in our experience, is not negligible.
To achieve full online anomaly detection capabilities, approaches frequently exploit statistical techniques that provide efficient training and analysis capabilites including changepoint detection~\cite{Basseville:1993}, Holt-Winters method and its extensions~\cite{Szmit:2012}, eccentricity analysis~\cite{Angelov:2014} 




This body of work addresses the problem of efficiently detecting anomalies, but contrarily to ADaaS they do not consider the problem of controlling the deployment and un-deployment of the anomaly detectors. Indeed, ADaaS can incorporate different kinds of anomaly detectors, including statistical anomaly detectors as reported in this paper, whose deployment can be fully controlled by the operator. Interestingly, the operator can change at any time the set of detected anomalies, based on the emerging needs and changes in the monitored data.  Even if a comprehensive anomaly detection as-a-service remains an open challenge~\cite{Yao:2017:ADS}, the degree of automation presented in ADaaS is a key enabling factor for its realization.

\section{Conclusion}
\label{sec:conclusion}

Cloud systems are large and complex software systems whose behavior can be hardly predicted and controlled. For instance, cloud applications can be dynamically scaled up and down based on workloads, while components can be updated based on emerging requirements. Although cloud systems can be monitored quite effectively despite changes, the anomaly detection logic is still quite expensive to modify and adapt to changes that may modify the structure and behavior of the observed system. 

In this paper we discussed how to apply the as-a-service paradigm to the anomaly detection logic, achieving Anomaly Detection as-a-service (ADaaS). We also propose an architecture that supports the ADaaS paradigm and that can work jointly with any monitoring system that stores data in time-series databases. 

We early experimented ADaaS with the Clearwater cloud system obtaining results that demonstrate how the ADaaS paradigm can be effectively used to handle the anomaly detection logic. Our future work includes experimenting the approach with a larger set of use cases to stress the flexibility, generality, and efficiency of the solution. 

\section*{Acknowledgements}
	This work has been supported by the H2020 5G-PPP Phase2 NGPaaS project (Grant Agreement No. 761557), by the H2020 ERC CoG Learn project (Grant Agreement No. 646867) and by the ``GAUSS'' national research project, which has been funded 	by the MIUR under the PRIN 2015 program (Contract 2015KWREMX).

\balance
\small
\bibliographystyle{abbrv}
\bibliography{bibliography,webreferences,tools}

\begin{thebibliography}{10}

\bibitem{aceto2013monitoringsurvey}
G.~Aceto, A.~Botta, W.~de~Donato, and A.~Pescapè.
\newblock {Cloud monitoring: A survey}.
\newblock {\em Computer Networks}, 57(9):2093--2115, 2013.

\bibitem{AHMAD2017134}
S.~Ahmad, A.~Lavin, S.~Purdy, and Z.~Agha.
\newblock Unsupervised real-time anomaly detection for streaming data.
\newblock {\em Neurocomputing}, 262:134 -- 147, 2017.
\newblock Online Real-Time Learning Strategies for Data Streams.

\bibitem{AKOUEMO2016948}
H.~N. Akouemo and R.~J. Povinelli.
\newblock Probabilistic anomaly detection in natural gas time series data.
\newblock {\em International Journal of Forecasting}, 32(3):948 -- 956, 2016.

\bibitem{amazon_2019_cloudwatch}
{Amazon Web Services, Inc.}
\newblock {CloudWatch}.
\newblock \url{https://aws.amazon.com/it/cloudwatch/}, 2019.
\newblock [Online; accessed 15-May-2019].

\bibitem{Angelov:2014}
P.~{Angelov}.
\newblock Anomaly detection based on eccentricity analysis.
\newblock In {\em 2014 IEEE Symposium on Evolving and Autonomous Learning
  Systems (EALS)}, pages 1--8, Dec 2014.

\bibitem{Basseville:1993}
M.~Basseville and I.~V. Nikiforov.
\newblock {\em Detection of Abrupt Changes: Theory and Application}.
\newblock Prentice-Hall, Inc., Upper Saddle River, NJ, USA, 1993.

\bibitem{buyya2009cloud}
R.~Buyya, C.~S. Yeo, S.~Venugopal, J.~Broberg, and I.~Brandic.
\newblock Cloud computing and emerging it platforms: Vision, hype, and reality
  for delivering computing as the 5th utility.
\newblock {\em Future Generation computer systems}, 25(6):599--616, 2009.

\bibitem{calero_monpaas:_2015}
J.~A. Calero and J.~G. Aguado.
\newblock {MonPaaS}: an adaptive monitoring platform as a service for cloud
  computing infrastructures and services.
\newblock {\em IEEE Transactions on Services Computing}, 8(1):65--78, 2015.

\bibitem{chandola_anomaly_2009}
V.~Chandola, A.~Banerjee, and V.~Kumar.
\newblock Anomaly detection: {A} survey.
\newblock {\em ACM Computing Surveys}, 41(3):1--58, July 2009.

\bibitem{Chandola:2008}
V.~{Chandola}, V.~{Mithal}, and V.~{Kumar}.
\newblock Comparative evaluation of anomaly detection techniques for sequence
  data.
\newblock In {\em 2008 Eighth IEEE International Conference on Data Mining},
  pages 743--748, Dec 2008.

\bibitem{chang2011libsvm}
C.-C. Chang and C.-J. Lin.
\newblock Libsvm: A library for support vector machines.
\newblock {\em ACM transactions on intelligent systems and technology (TIST)},
  2(3):27, 2011.

\bibitem{Chen:2015}
P.~{Chen}, S.~{Yang}, and J.~A. {McCann}.
\newblock Distributed real-time anomaly detection in networked industrial
  sensing systems.
\newblock {\em IEEE Transactions on Industrial Electronics}, 62(6):3832--3842,
  June 2015.

\bibitem{Dean:UBL:ICAC:2012}
D.~J. Dean, H.~Nguyen, and X.~Gu.
\newblock Ubl: Unsupervised behavior learning for predicting performance
  anomalies in virtualized cloud systems.
\newblock ICAC '12, pages 191--200, 2012.

\bibitem{elastic_2019_elasticsearch}
{Elasticsearch BV}.
\newblock {Elasticsearch: RESTful, Distributed Search \& Analytics}.
\newblock \url{ https://www.elastic.co/products/elasticsearch }, 2019.
\newblock [Online; accessed 15-May-2019].

\bibitem{gartner_forecast_21.4}
{Gartner, Inc.}
\newblock {Gartner Forecasts Worldwide Public Cloud Revenue to Grow 21.4
  Percent in 2018}.
\newblock \url{ https://www.gartner.com/newsroom/id/3871416 }, 2018.
\newblock [Online; accessed 15-May-2019].

\bibitem{hp_2019_monasca}
{Hewlett-Packard Enterprise Development LP}.
\newblock {Monasca - an {OpenStack} {Community} project}.
\newblock \url{http://http://monasca.io/ }, 2017.
\newblock [Online; accessed 15-May-2019].

\bibitem{josep2010view}
A.~D. JoSEP, R.~KAtz, A.~KonWinSKi, L.~Gunho, D.~PAttERSon, and A.~RABKin.
\newblock A view of cloud computing.
\newblock {\em Communications of the ACM}, 53(4), 2010.

\bibitem{keogh:2005}
E.~{Keogh}, J.~{Lin}, and A.~{Fu}.
\newblock Hot sax: efficiently finding the most unusual time series
  subsequence.
\newblock In {\em Fifth IEEE International Conference on Data Mining
  (ICDM'05)}, pages 8 pp.--, Nov 2005.

\bibitem{Laptev:2015:GSF:2783258.2788611}
N.~Laptev, S.~Amizadeh, and I.~Flint.
\newblock Generic and scalable framework for automated time-series anomaly
  detection.
\newblock In {\em Proceedings of the 21th ACM SIGKDD International Conference
  on Knowledge Discovery and Data Mining}, KDD '15, pages 1939--1947, New York,
  NY, USA, 2015. ACM.

\bibitem{Mariani:ICST:18}
L.~{Mariani}, C.~{Monni}, M.~{Pezzé}, O.~{Riganelli}, and R.~{Xin}.
\newblock Localizing faults in cloud systems.
\newblock In {\em Proceedings of the International Conference on Software
  Testing, Verification and Validation (ICST)}, pages 262--273, 2018.

\bibitem{clearwater}
{Metaswitch Networks}.
\newblock {Project ClearWater}.
\newblock \url{ https://www.projectclearwater.org/}.
\newblock [Online; accessed 21-Jul-2019].

\bibitem{surus:2015}
{Netflix}.
\newblock {Surus}.
\newblock \url{ https://github.com/Netflix/Surus}, 2015.
\newblock [Online; accessed 20-Jul-2019].

\bibitem{orru2018chmmiot}
M.~Orr\'{u}, M.~Mobilio, A.~Shatnawi, O.~Riganelli, A.~Tundo, and L.~Mariani.
\newblock {Model-Based Monitoring for IoTs Smart Cities Applications}.
\newblock In {\em 4th Italian Conference on ICT for Smart Cities And
  Communities i-CiTies 2018}, 2018.

\bibitem{prometheus}
{Prometheus Authors}.
\newblock Prometheus.
\newblock \url{https://prometheus.io/}, 2019.
\newblock [Online; accessed 15-May-2019].

\bibitem{sabahi_cloud_2011}
F.~{Sabahi}.
\newblock Cloud computing security threats and responses.
\newblock In {\em 2011 IEEE 3rd International Conference on Communication
  Software and Networks}, pages 245--249, May 2011.

\bibitem{Sauvanaud:Anomaly:ISSRE:2016}
C.~Sauvanaud, K.~Lazri, M.~Kaaniche, and K.~Kanoun.
\newblock Anomaly detection and root cause localization in virtual network
  functions.
\newblock ISSRE '16, 2016.

\bibitem{shatnawi2018chmm}
A.~Shatnawi, M.~Orr\'{u}, M.~Mobilio, O.~Riganelli, and L.~Mariani.
\newblock {CloudHealth: {A} Model-Driven Approach to Watch the Health of Cloud
  Services}.
\newblock In {\em Proceedings of the 1st International Workshop on Software
  Health (SoHeal 2018)}, pages 40--47. ACM/IEEE, 2018.

\bibitem{Spinosa:2007}
E.~J. Spinosa, A.~P. de~Leon F.~de Carvalho, and J.~a. Gama.
\newblock Olindda: A cluster-based approach for detecting novelty and concept
  drift in data streams.
\newblock In {\em Proceedings of the 2007 ACM Symposium on Applied Computing},
  SAC '07, pages 448--452, New York, NY, USA, 2007. ACM.

\bibitem{Szmit:2012}
M.~Szmit and A.~Szmit.
\newblock Usage of modified holt-winters method in the anomaly detection of
  network traffic: Case studies.
\newblock {\em Journal of Computer Networks and Communications}, 2012, 2012.

\bibitem{tundo:2019}
A.~Tundo, M.~Mobilio, M.~Orr\'{u}, O.~Riganelli, M.~Guzm\'{a}n, and L.~Mariani.
\newblock {VARYS}: An agnostic model-driven monitoring-as-a-service framework
  for the cloud.
\newblock In {\em Proceedings to the 27th ACM Joint European Software
  Engineering Conference and Symposium on the Foundations of Software
  Engineering Tallinn, Estonia, 26-30}, Aug. 2019.

\bibitem{vaquero_dynamically_2011}
L.~M. Vaquero, L.~Rodero-Merino, and R.~Buyya.
\newblock Dynamically {Scaling} {Applications} in the {Cloud}.
\newblock {\em SIGCOMM Comput. Commun. Rev.}, 41(1):45--52, Jan. 2011.

\bibitem{Yao:2017:ADS}
D.~Yao, X.~Shu, L.~Cheng, S.~J. Stolfo, and E.~Bertino.
\newblock {\em Anomaly Detection As a Service: Challenges, Advances, and
  Opportunities}.
\newblock Morgan \& Claypool Publishers, 2017.

\bibitem{yazir_dynamic_2010}
Y.~O. Yazir, C.~Matthews, R.~Farahbod, S.~Neville, A.~Guitouni, S.~Ganti, and
  Y.~Coady.
\newblock Dynamic {Resource} {Allocation} in {Computing} {Clouds} {Using}
  {Distributed} {Multiple} {Criteria} {Decision} {Analysis}.
\newblock In {\em 2010 {IEEE} 3rd {International} {Conference} on {Cloud}
  {Computing}}, pages 91--98, July 2010.

\bibitem{Tan:anomalyPrediction:ICDCS:2012}
T.~Yongmin, N.~Hiep, S.~Zhiming, G.~Xiaohui, V.~Chitra, and R.~Deepak.
\newblock Prepare: Predictive performance anomaly prevention for virtualized
  cloud systems.
\newblock ICDCS '12, pages 285--294, 2012.

\end{thebibliography}

\end{document}